\documentclass{PoS}

\usepackage{graphicx}
\usepackage{epsfig}
\newcommand{\be}{\begin{equation}}
\newcommand{\ee}{\end{equation}}
\newcommand{\beq}{\begin{eqnarray}}
\newcommand{\eeq}{\end{eqnarray}}

\title{Nucleon  and Nucleon-to-$\Delta$ Axial  Form Factors from 
Lattice QCD}

\ShortTitle{Nucleon  and Nucleon-to-$\Delta$ Axial  Form Factors from 
Lattice QCD}

\author{Constantia Alexandrou\\
Department of Physics, University of Cyprus, CY-1678 Nicosia, Cyprus\\
E-mail: \email{alexand@ucy.ac.cy}}
\author{Giannis Koutsou\\
Department of Physics, University of Cyprus, CY-1678 Nicosia, Cyprus\\
E-mail: \email{koutsou@ucy.ac.cy}}
\author{Theodoros Leontiou\\
Department of Physics, University of Cyprus, CY-1678 Nicosia, Cyprus\\
E-mail: \email{t.leontiou@ucy.ac.cy}}
\author{John W. Negele\\
        Center for Theoretical Physics, Laboratory for
 Nuclear Science and Department of Physics, Massachusetts Institute of
Technology, Cambridge, Massachusetts 02139, U.S.A.\\
        E-mail: \email{negele@mit.edu}}
\author{\speaker{Antonios Tsapalis}\\
        University of Athens, Institute of Accelerating Systems
    and Applications, Athens, Greece \\
        E-mail: \email{a.tsapalis@iasa.gr}}

\abstract{
We present results on the 
 nucleon axial vector form factors 
$G_A(q^2)$ and $G_p(q^2)$ in the quenched theory
and using two degenerate 
flavors of dynamical Wilson fermions for momentum transfer squared from
about $0.1$ to about $2$~GeV$^2$ and for pion masses in the range
of 380 to 600 MeV. 
We also present results on the corresponding $N$ to $\Delta$ 
axial vector transition form factors $C_5^A(q^2)$ and $C_6^A(q^2)$
using, in addition to  Wilson fermions, domain wall valence quarks
and dynamical  staggered sea quarks provided by the MILC
collaboration.

}

\FullConference{The XXV International Symposium on Lattice Field Theory\\
                 July 30 - August 4 2007\\
                 Regensburg, Germany}

\begin{document}

\section{Introduction}
Form factors are fundamental quantities which
characterize important features of the hadrons such as their size, 
charge distribution and magnetization.
In this work we present a calculation of the axial form factors of the nucleon
and the nucleon (N) to $\Delta$  transition~\cite{PRDaxial}. 
Spontaneous breaking of axial  symmetry in QCD is manifest in 
the existence of light pseudo-Goldstone bosons. 
The smallness of the pion mass is connected to the axial
symmetry breaking through the {\it partial conservation of axial current}
(PCAC) hypothesis, which relates  the divergence 
of the axial vector current to  the pion field. 
This translates
into the pion-pole dominance of the
induced nucleon  pseudoscalar form factor, $G_p(q^2)$. 
Furthermore, it relates  the nucleon axial charge $g_A$, 
 to the $\pi NN$ coupling, $g_{\pi NN}$ via the well known Goldberger-Treiman
relation (GTR), $m_N g_A=f_\pi g_{\pi NN}$, where $f_\pi$ is the pion decay 
constant and $m_N$ the mass of the nucleon.
Similarly, PCAC applied to the N to $\Delta$ transition, leads to the
assumption of  pion-pole
dominance in the case of $C_6^A$ and to the non-diagonal  GTR, 
$2m_N C_5^A=f_\pi g_{\pi N\Delta}$, where $g_{\pi N\Delta}$ is the strong
coupling constant associated with the $\pi N\Delta$ vertex.
Our calculation of the nucleon axial vector form factors,  $G_A(q^2)$
and $G_p(q^2)$, as well as the pseudoscalar form factor $G_{\pi NN}$,
for a range of pion masses down to 380 MeV, both
in the quenched theory and using  two degenerate 
flavors of dynamical Wilson fermions,  enables us
to  check pion-pole dominance and the Goldberger-Treiman relation.
Similarly, in the case of  the N to $\Delta (1232)$ weak current transition,
we evaluate the corresponding 
dominant form factors $ C_5^A(q^2)$ and $C_6^A(q^2)$, as well as
the pseudoscalar form factor $G_{\pi N\Delta}$, using besides 
Wilson fermions, dynamical staggered quark configurations 
generated by the MILC 
collaboration and domain wall fermions for pion masses as low as 360~MeV.  
The evaluation of both nucleon and N to $\Delta$ transition form 
factors enables us to check the $q^2$-dependence of ratios of 
transition to nucleon form factors, such as 
$G_{\pi N\Delta}(q^2)/G_{\pi NN}(q^2)$
and $C_5^A(q^2)/G_A(q^2)$, which show a weaker
quark mass dependence and are more likely to
 show less  sensitivity to lattice
systematics.

\vspace*{-0.3cm}

\section{Axial Form Factors and the Pion Pole}

\vspace*{-0.1cm}
We take the u- and the d- quarks to be degenerate
and work 
in terms of the fermion isospin doublet $\psi$. The axial vector  
and pseudoscalar currents are given by
$A_{\mu}^a(x)= \bar{\psi}(x)\gamma_\mu \gamma_5\frac{\tau^a}{2}\psi(x)$ and 
$ P^a(x)= \bar{\psi}(x)\gamma_5 \frac{\tau^a}{2}\psi(x) $ respectively,
where $\tau^a$ are Pauli-matrices acting in flavor space.
 The matrix element  of the weak axial vector current between nucleon states
is written  as 
\small
\be
\langle N(p',s')|A_\mu^3|N(p,s)\rangle= i \Bigg(\frac{
            m_N^2}{E_N(p')E_N(p)}\Bigg)^{1/2} 
            \bar{u}(p',s') \Bigg[
            \left(G_A(q^2)\gamma_\mu\gamma_5 
            +\frac{q_\mu}{2m_N}G_p(q^2)\right)\Bigg]\frac{\tau^3}{2}u(p,s)
\label{NN axial}
\ee 
\normalsize
with the axial vector, $G_A(q^2)$, and the  induced pseudoscalar form factor,
  $G_p(q^2)$, being functions of the invariant 
momentum transfer squared, $q^2=(p^\prime-p)^2$.
Experimentally, the axial charge of the nucleon, $g_A \equiv G_A(0)$ 
is very well known from the neutron $\beta$-decay and takes the value
 $g_A=1.2695(29)$.
 The momentum dependence of $G_A$ has been extracted from 
pion electroproduction or quasielastic neutrino scattering 
experiments (see Ref.~\cite{experiment} for reviews) and 
it  is conventionally parameterized by a dipole form, 
$
G_A(q^2) = g_A/\left( 1-\frac{q^2}{M_A^2} \right)^2
$,
with an  axial mass, $M_A=1.026 \pm 0.0021\, {\rm GeV}$.
The induced pseudoscalar form factor $G_p(q^2)$ is less well studied
 experimentally~\cite{experiment}, with ordinary and radiative muon 
capture experiments 
giving different results. Both form factors have been studied in the 
context of chiral effective theories~\cite{experiment}.  
 
In QCD the explicit breaking of  axial current conservation 
leads to the axial Ward-Takahashi identity,
$\partial^\mu A_\mu^a=2m_q P^a$, where $m_q$ is the renormalized quark mass.
Using PCAC we have the relation,
$\partial^\mu A_\mu^a=f_\pi m_\pi^2 \pi^a $,
and therefore the pion field is expressed in terms of the pseudoscalar
current as
$\pi^a=2 m_q P^a/f_\pi m_\pi^2$.
 Taking the matrix element of the pseudoscalar density between 
nucleon states,  the $\pi NN$ form factor is obtained via
\be
 2m_q<N(p^\prime,s^\prime)|P^3|N(p,s)>= 
\Bigg(\frac{m_N^2}{E_N({\bf p}')E_N({\bf p})}\Bigg)^{1/2} 
\frac{f_\pi m_\pi^2\> G_{\pi NN}(q^2)}
{m_\pi^2-q^2}\>\bar{u}(p^\prime,s^\prime)i\gamma_5 \frac{\tau^3}{2}u(p,s).
\label{g_piNN}
\ee
Using the PCAC hypothesis
together with Eq.~(\ref{g_piNN}) we obtain
the diagonal Goldberger-Treiman relation 
\beq 
 G_A(q^2)+\frac{q^2}{4m_N^2} G_p(q^2) = 
\frac{1}{2m_N}\frac{2G_{\pi N N}(q^2)f_\pi m_\pi^2}{m_\pi^2-q^2} \quad.  
\label{GTR}
\eeq
In the chiral limit, the pole on the right hand size 
of Eq.~(\ref{GTR}) must be 
compensated by a similar singularity in  $G_p(q^2)$ 
since $G_A(0)$ is finite.
 Therefore, assuming  pion-pole dominance, $G_p(q^2)$ is given 
by
\beq 
\frac{1}{2m_N}G_p(q^2)&\sim &\frac{2G_{\pi NN}(q^2) f_\pi}{m_\pi^2-q^2} 
\label{GP}
\eeq
and substituting in Eq.~(\ref{GTR})  we obtain the well known relation,
$m_N G_A(q^2) = f_\pi G_{\pi NN}(q^2)$.

The invariant proton to $\Delta^+$ weak matrix element is expressed
in terms of four transition
form factors as
\beq
<\Delta(p^{\prime},s^\prime)|A^3_{\mu}|N(p,s)> &=& i\sqrt{\frac{2}{3}} 
\left(\frac{m_\Delta m_N}{E_\Delta({\bf p}^\prime) E_N({\bf p})}\right)^{1/2}
\bar{u}_{\Delta^+}^\lambda(p^\prime,s^\prime)\nonumber \\
&\>&\hspace*{-5cm}
\biggl[\left (\frac{C^A_3(q^2)}{m_N}\gamma^\nu + \frac{C^A_4(q^2)}{m^2_N}p{^{\prime \nu}}\right)  
\left(g_{\lambda\mu}g_{\rho\nu}-g_{\lambda\rho}g_{\mu\nu}\right)q^\rho
+C^A_5(q^2) g_{\lambda\mu} +\frac{C^A_6(q^2)}{m^2_N} q_\lambda q_\mu \biggr]
u_P(p,s).
\label{ND axial}
\eeq
The form factors
$C^A_3(q^2)$ and $C^A_4(q^2)$ belong to the transverse part of the axial 
current and, as was recently demonstrated~\cite{PRLaxial},
both are suppressed relative to $C^A_5(q^2)$ and   $C^A_6(q^2)$. 
 The pseudoscalar transition form factor, $G_{\pi N\Delta}$, is defined 
similarly to $G_{\pi NN}$, via 
\small
\be 
 2m_q<\Delta(p^\prime,s^\prime)|P^3|N(p,s)> = i\sqrt{\frac{2}{3}}
\left(\frac{m_\Delta m_N}{E_\Delta({\bf p}^\prime) E_N({\bf p})}\right)^{1/2}
\frac{f_\pi m_\pi^2 \>G_{\pi N\Delta}(q^2)}
{m_\pi^2-q^2}
\bar{u}_{\Delta^+}^\nu(p^\prime,s^\prime)\frac{q_\nu}{2m_N} u_P(p,s)
\label{g_piND} \quad.
\ee
\normalsize
The $\pi NN$ and $\pi N\Delta$ coupling constants are defined at
$q^2=m_\pi^2$ as $g_{\pi NN}=G_{\pi NN}(m_{\pi}^2)$ and
$g_{\pi N\Delta}=G_{\pi N\Delta}(m_{\pi}^2)$.
The dominant form factors $C^A_5(q^2)$ and $C^A_6(q^2)$, which belong 
to the longitudinal 
part of the axial current, are related via PCAC to $G_{\pi N\Delta}(q^2)$: 
\be
 C_5^A(q^2)+\frac{q^2}{m_N^2} C_6^A(q^2) = 
\frac{1}{2m_N}\frac{G_{\pi N \Delta}(q^2)f_\pi m_\pi^2}{m_\pi^2-q^2} \quad,
\label{GTR_ND}
\ee
which is known as the non-diagonal GTR.
Pion-pole dominance for $C_6^A(q^2)$, as for $G_p(q^2)$,  
leads to the relations
$
C_6^A(q^2)\sim m_N G_{\pi N\Delta}(q^2) f_\pi/ 2(m_\pi^2-q^2)$ and 
$2m_N C_5^A(q^2) = f_\pi G_{\pi N\Delta}(q^2)$.
It is clear from the above relations that $C^A_5$ is analogous 
to $G_A$ and $C^A_6$ 
analogous to $G_p$.

\section{Lattice techniques}
The techniques used for the evaluation of the axial form factors were
developed for the study of the electromagnetic form factors~\cite{NN lattice,PRL_quenched, ND dynamical}. In the evaluation of the three-point functions
involved in the calculation of the electromagnetic nucleon and N to $\Delta$
transition form factors we use sequential inversion through the sink, thereby
obtaining the form factors for all momentum transfers. This method
requires fixing the source-sink time separation, as well
as the initial and
final hadron states, but allows the insertion of any operator
with arbitrary momentum at any
time slice. 
An important ingredient of our method is the construction
of optimal sources and sinks
using a linear combination of  interpolating fields. 
It was shown in Refs.~\cite{PRDaxial,PRLaxial}
that, for axial operators,  the most symmetric
linear combination of  matrix elements that can be considered is 
\small
\be 
 \sum_{k=1}^3\Pi^A ({\bf 0}, -{\bf q}\; ;
\Gamma_k ;\mu = j) = 
i \; \frac{C}{4m_N}\biggl[(E_N+m_N)
\left( \delta_{1,j} + \delta_{2,j} + \delta_{3,j} \right) G_A(Q^2)
- (q_1 + q_2 + q_3) \frac{q_j}{2 m_N} G_p(Q^2) \biggr], 
\label{SA optimal}
\ee
\normalsize
where $j = 1,2,3$ labels the spatial current direction, $\Gamma$ projects to
definite nucleon spin states and $C$ is a kinematical factor.
 $\Pi^A ({\bf p'}, {\bf p}\; ;\Gamma ;\mu)$ is the large Euclidean
time limit of an appropriately constructed  ratio of the three-point
function to two-point functions, in which time dependencies of the
time evolution and overlaps of the initial 
trial state  and the nucleon state cancel.
Since this optimized sink, involving momentum in all spatial directions,
coincides with the one   calculated for the electromagnetic 
current
{\it no new sequential inversions} are required for the axial vector current.
The same holds for the pseudoscalar current and
the N to $\Delta$ optimized sinks.
As indicated in Eq.~(\ref{SA optimal}), we use kinematics 
where the final nucleon
 is produced at rest
and therefore the momentum transfer 
${\bf q}={\bf p}^{\prime}-{\bf p}=-{\bf p}$.
We take
$-q^2=Q^2>0$
with $Q^2$ being the Euclidean momentum transfer squared.

\begin{table}[h]\begin{center}
\vspace*{-0.15cm}
\begin{tabular}{|c|c|c|c|c|}
\hline
\multicolumn{1}{|c|}{\# confs } &
\multicolumn{1}{ c|}{$\kappa$ or $am_l$ } &
\multicolumn{1}{ c|}{$ m_\pi$ (GeV)} &
\multicolumn{1}{ c|}{$ M_N$ (GeV)} &
\multicolumn{1}{ c|}{$ M_{\Delta}$ (GeV) }
\\
\hline
\multicolumn{5}{|c|}{Quenched $32^3\times 64$ \hspace*{0.5cm} $a^{-1}=2.14(6)$ GeV}
 \\ \hline
  200            &  0.1554 & 0.563(4) &  1.267(11)   & 1.470(15) \\
  200            &  0.1558 & 0.490(4) &  1.190(13)   & 1.425(16) \\
  200            &  0.1562 & 0.411(4) &  1.109(13)   & 1.382(19)\\
    &  $\kappa_c=$0.1571   & 0.       &  0.938(9)  &            \\
\hline
\multicolumn{5}{|c|}{Unquenched Wilson $24^3\times 40 $~\cite{TxL}  \hspace*{0.5cm}
$a^{-1}=2.56(10)$ GeV}
 \\\hline
 185                &  0.1575  & 0.691(8) & 1.485(18) & 1.687(15)\\
 157                &  0.1580  & 0.509(8) & 1.280(26) & 1.559(19) \\
\hline
\multicolumn{5}{|c|}{Unquenched Wilson $24^3\times 32 $~\cite{Urbach}  \hspace*{0.5cm}
$a^{-1}=2.56(10)$ GeV}
\\\hline
 200                &  0.15825 & 0.384(8) & 1.083(18)  & 1.395(18)  \\
                    & $\kappa_c=0.1585$& 0. & 0.938(33)&           \\
\hline
\multicolumn{5}{|c|}{MILC $20^3\times 64 $  \hspace*{0.5cm}
$a^{-1}=1.58$ GeV}
 \\\hline
 200                &  0.03  & 0.594(1) & 1.416(20)&  1.683(22)\\
 198                &  0.02  & 0.498(3) & 1.261(17)&  1.589(35)\\
 100                &  0.01  & 0.362(5) & 1.139(25)&  1.488(71)\\
\hline
\multicolumn{5}{|c|}{MILC $28^3\times 64 $  \hspace*{0.5cm}
$a^{-1}=1.58$ GeV}
\\\hline
 150                &  0.01 & 0.357(2) & 1.210(24)  & 1.514(41) \\
\hline
\end{tabular}
\end{center}
\vspace*{-0.5cm}\caption{The number of configurations,
 the hopping parameter, $\kappa$, for the
case of  Wilson fermions or the mass of the light quarks, $m_l$,
for the MILC staggered quarks, and
 the pion,  nucleon and $\Delta$ masses.} 
\label{table:lattices}
\end{table}
In Table I we collect the parameters for our calculation. In 
the so called hybrid approach, that uses  staggered sea quarks and
 domain wall valence
quarks,
 we have no ${\cal O}(a)$ artifacts, 
unlike Wilson fermions where cutoff effects
are ${\cal O}(a)$. We use the same parameters for the domain wall operator 
as those used in Ref.~\cite{LHPC}, namely we take
the length of the fifth dimension, $L_5/a=16$ 
and the valence quark mass that is tuned 
to reproduce the pion mass 
 calculated with the staggered quark action. 
Finite volume effects can be assessed by comparing results at the
lowest pion mass. It was shown~\cite{PRDaxial} that results on the $20^3$
spatial volume were consistent with the results on the $28^3$ lattice.
  The source-sink separation is optimized so that,
on the one hand, it is sufficiently large to ensure
  ground state dominance and, on the other, small enough
so that  gauge noise is kept at a minimum. 
 In all cases, we use the non-perturbatively determined value
for the axial renormalization constant,  $Z_A$. Note that the
pseudoscalar renormalization constant is not needed for the
quantities under consideration here.

\section{Results}

We first consider the ratios $G_{\pi N\Delta}(Q^2)/G_{\pi NN}(Q^2)$
and  $2C_5^A(Q^2)/G_A(Q^2)$. We note that, in the ratio 
$G_{\pi N\Delta}(Q^2)/G_{\pi NN}(Q^2)$,
the renormalized quark mass cancels
 eliminating one source of systematic error.

\begin{figure}[h]
 \begin{minipage}[h]{7.1cm}
\epsfxsize=8truecm \epsfysize=5truecm
{\mbox{\includegraphics[height=6cm,width=7.5cm]{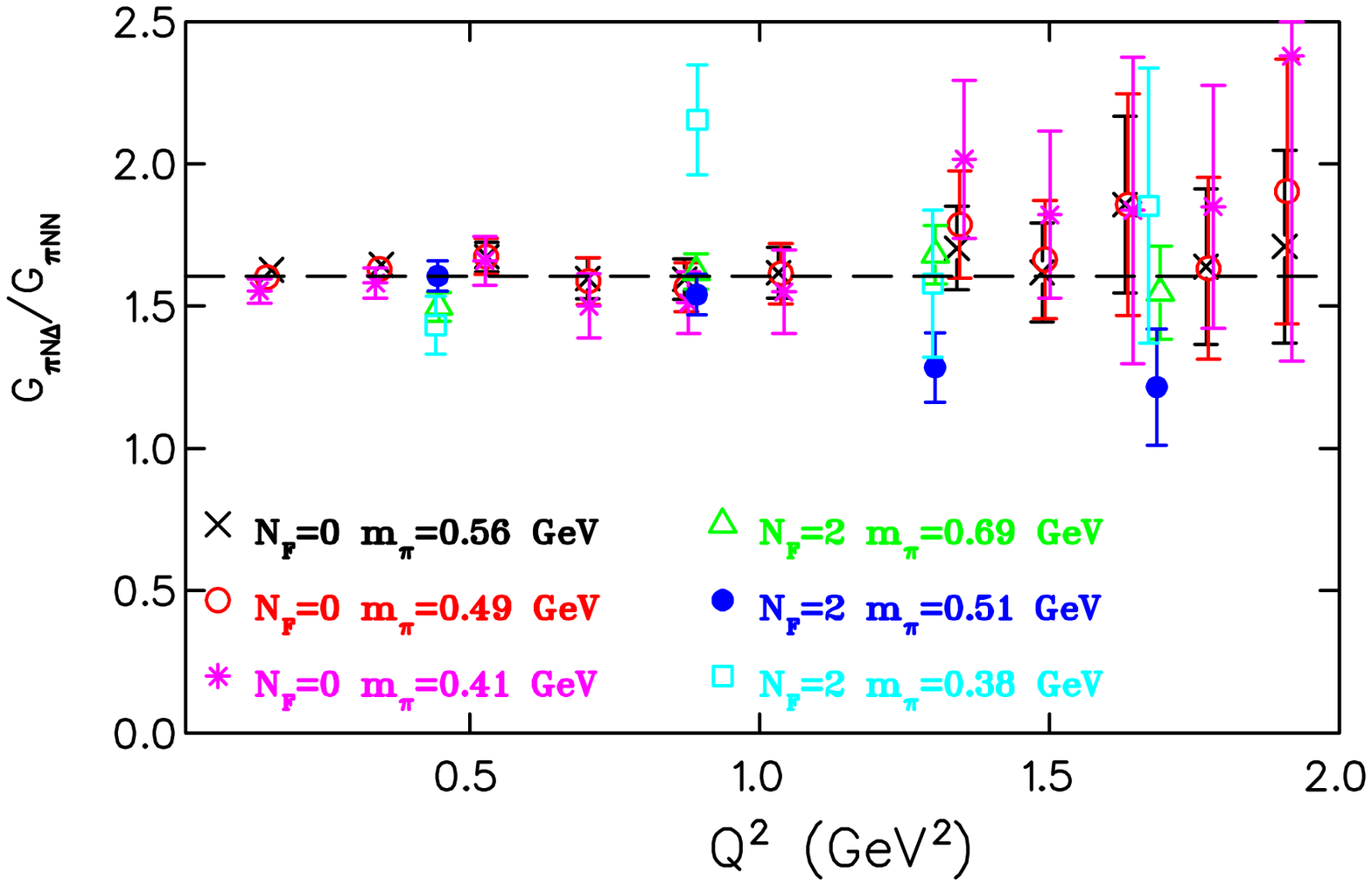}}}
\caption{The ratio of form factors $G_{\pi N\Delta}(Q^2)/G_{\pi NN}(Q^2)$
 as a function of $Q^2$ for
Wilson fermions for the quenched theory, denoted by $N_F=0$, 
and for two dynamical Wilson quarks, denoted
by $N_F=2$. 
}
\label{fig:gpiNDovergpiNN}
\end{minipage}\hspace*{0.4cm}
    \begin{minipage}[h]{7.1cm}\vspace*{-0.7cm}
{\mbox{\includegraphics[height=6cm,width=7.5cm]{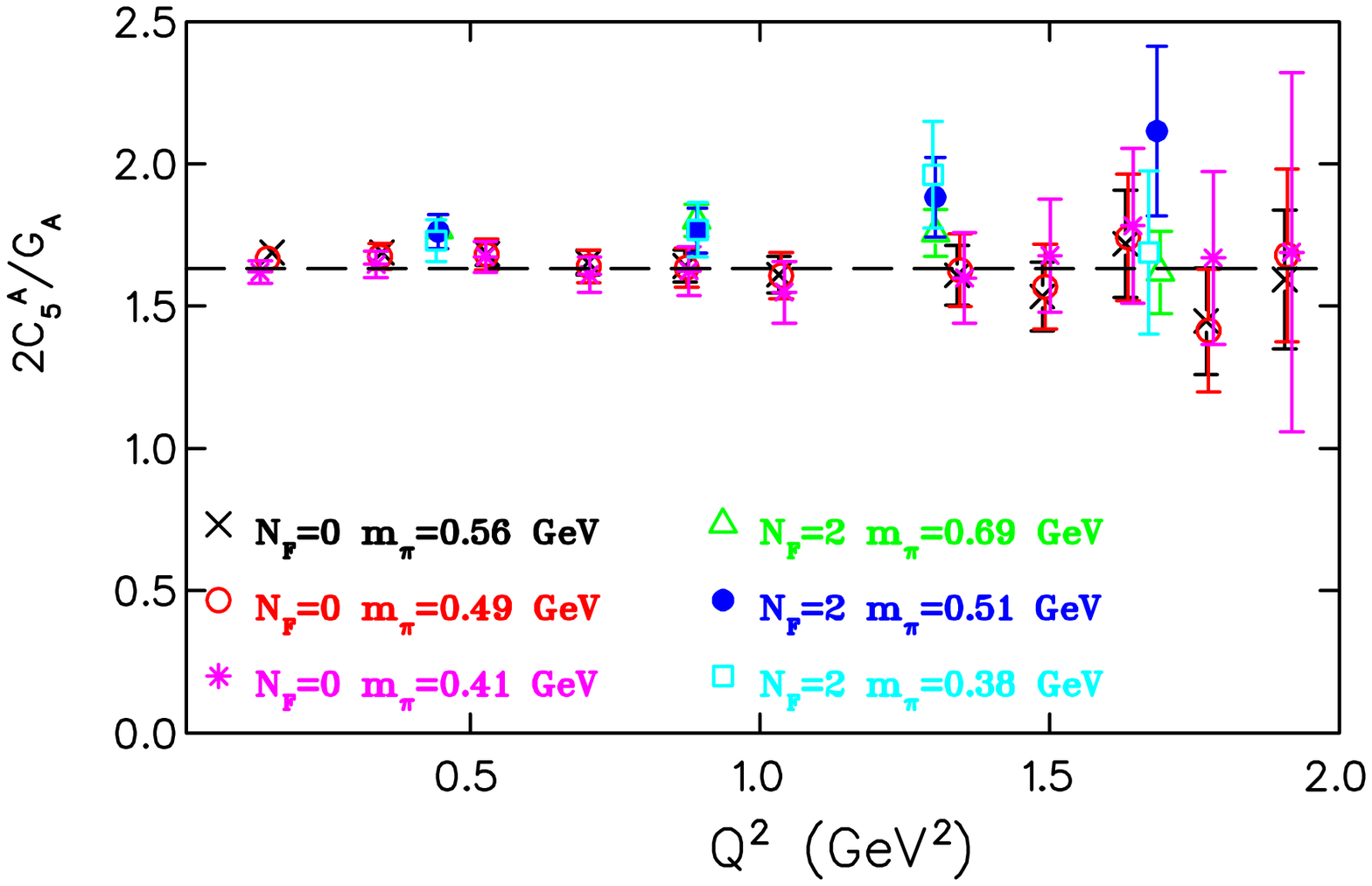}}}
\caption{The ratio of $2C_5^A(Q^2)/G_A(Q^2)$ as a function of $Q^2$. 
The notation is the same as that of Fig.~1.} 
\label{fig:CA5overGA}
\end{minipage}
\end{figure}
As can be seen in Figs.~\ref{fig:gpiNDovergpiNN}
and \ref{fig:CA5overGA}, 
these ratios show weak
dependence on the quark mass and are therefore
 more suited for comparison
with physical results.
Both these ratios are $Q^2$ independent and can be thus
fitted to a constant.
Fitting  the quenched data, which carry the smallest statistical errors,
 we obtain the value
of 1.60(2) shown by the dashed line.
Taking the ratio of the diagonal and non-diagonal GTRs 
$m_N G_A=f_\pi G_{\pi NN}$ and  $2m_N C_5^A=f_\pi G_{\pi N\Delta}$,
we find that 
 $G_{\pi N\Delta}(Q^2)/G_{\pi NN}(Q^2)=2C_5^A(Q^2)/G_A(Q^2)$. 
In Fig.~\ref{fig:CA5overGA}, we show the ratio $2C_5^A(Q^2)/G_A(Q^2)$,
 which is indeed
also $Q^2$ independent, and fitting to the quenched
data we find  the value of $1.63(1)$ shown by the dashed line.
They are also approximately equal in the unquenched case~\cite{PRDaxial}. 
Therefore,  on the level of ratios, the GTRs are
satisfied.

\begin{figure}[h]
 \begin{minipage}[h]{7.5cm}
{\mbox{\includegraphics[height=7.6cm,width=7.5cm]{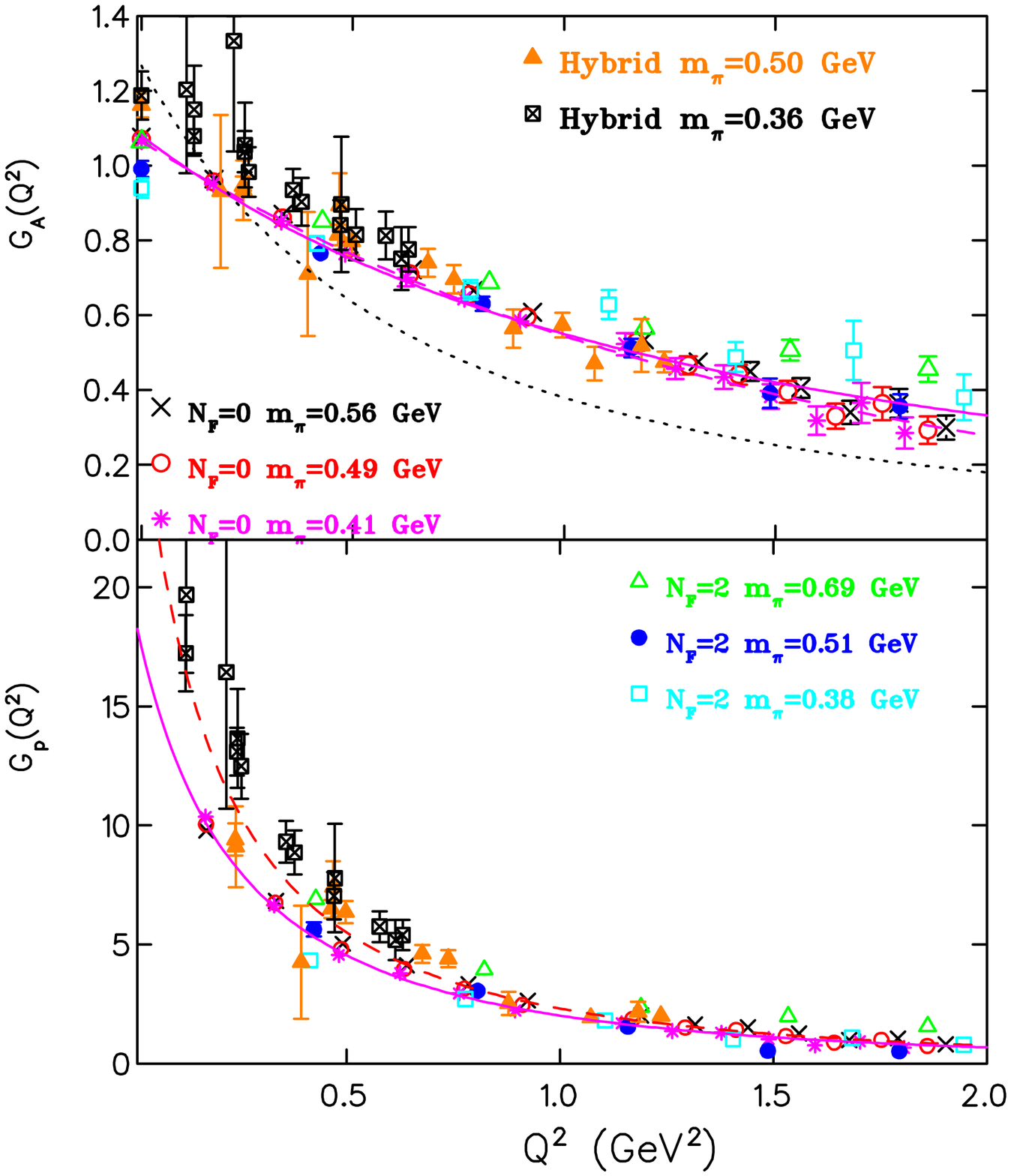}}}
    \caption{ $G_A(Q^2)$
(upper) and $G_p(Q^2)$(lower) for the quenched theory  and 
for two dynamical Wilson fermions. Results in the hybrid
approach are from~\cite{LHPC}. 
}
\label{fig:GAHA} 
    \end{minipage}
    \hfill
    \begin{minipage}[h]{7.1cm}\hspace*{-0.4cm}
{\mbox{\includegraphics[height=7.6cm,width=7.5cm]{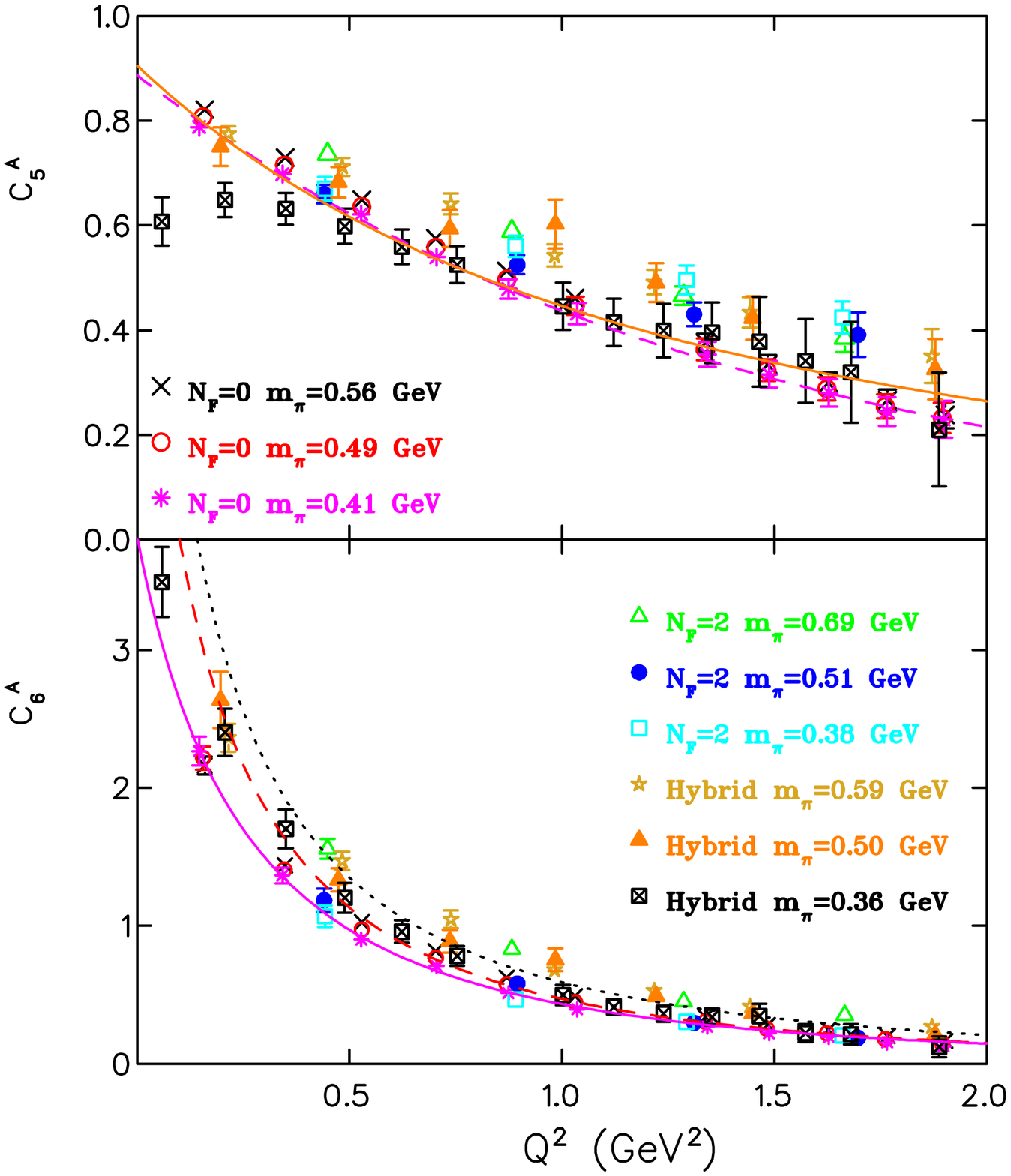}}}
    \caption{ $C_5^A(Q^2)$ (upper) and $C_6^A(Q^2)$ (lower) 
for  Wilson fermions and in the hybrid approach.
$\hspace*{5cm}$
}
\label{fig:CA5CA6}
    \end{minipage}
\vspace*{-0.2cm}
\end{figure}

In  Fig.~\ref{fig:GAHA}, we show our
results for the nucleon axial form factors using Wilson fermions.
We 
also include recent  results obtained in the hybrid approach~\cite{LHPC}
using similar parameters to those  used in our evaluation
of the N to $\Delta$ axial form factors. 
The main observation is  that, at the smallest
domain wall quark mass, results in the hybrid approach deviate from
quenched results.
 In particular, we
note that the value of the  nucleon axial charge $g_A$
 becomes larger in the hybrid scheme
approaching the experimental value. The dotted line shows the 
dipole fit to the experimental data. As can be seen, lattice
results fall off slower than experiment. However, the
large deviations observed at low $Q^2$ for a pion mass of about 360~MeV
point to large pion cloud effects, which tend to increase the form factors
at small $Q^2$. The solid line  is a dipole fit of the quenched 
$G_A(Q^2)$ at a pion mass of $411$~MeV, yielding an axial mass 
$M_A=1.58(3)$~GeV.
 We note that a fit to an exponential of the form 
$\tilde{g}_0\, {\rm exp}(-Q^2/\tilde{m}_A^2)$, also describes
the lattice data,  
yielding a curve that is  indistinguishable from the dipole fit.

Very similar behavior is observed for the transition form factor $C_5^A (Q^2)$
 shown in Fig.~\ref{fig:CA5CA6}, where a dipole fit, shown with the solid line,
provides a good description of the lattice results. Again the axial
mass obtained by fitting the quenched results at the lowest mass is larger
than  the value of $1.28(10)$~GeV  extracted from available experimental 
results~\cite{Kitagaki}. A fit to an  exponential form also provides
 a good fit to the lattice data.


Having fitted $G_A(Q^2)$ and $C_5^A(Q^2)$, 
the $Q^2$-dependence for the
 form factors $G_p(Q^2)$ and $C_6^A(Q^2)$ can be obtained using 
pion-pole dominance. The resulting curves are shown by the dashed lines
in Figs.~\ref{fig:GAHA} and \ref{fig:CA5CA6} for the quenched theory
at the lowest 
pion mass 
 and show deviations at low $Q^2$. The dotted line is the
corresponding result in the hybrid approach at the smallest pion mass.
In addition we 
show curves that are obtained using
\vspace*{-0.3cm}
\be
\hspace*{1.3cm}\frac{G_p(Q^2)}{G_A(Q^2)}=\frac{g_0}{(Q^2/m^2+1)} 
\label{fit functions}
\ee
with a corresponding expression for
$C_6^A(Q^2)/C_5^A(Q^2)$, where $g_0$ and $m$ are treated as fit parameters.
As expected, this provides a
good description of the $Q^2$-dependence for $G_p(Q^2)$  and $C_6^A(Q^2)$
as
shown by the solid lines, which correspond to the fits of  the quenched data at
 the lowest pion mass.

\section{Conclusions}
We present results for the nucleon axial vector form factors 
$G_A(Q^2)$ and
$G_p(Q^2)$, as well as for the corresponding 
N to $\Delta$ axial transition form factors $C_5^A(Q^2)$ and $C_6^A(Q^2)$. 
The $\pi NN$
and $\pi N\Delta$ form factors $G_{\pi NN}(Q^2)$ and $G_{\pi N \Delta}(Q^2)$
are also evaluated. 
One of the main conclusions is that  $G_{\pi NN}$ and $G_{\pi N \Delta}$
have  the same  $Q^2$ dependence yielding a ratio of 
$G_{\pi N\Delta}(Q^2)/ G_{\pi NN}(Q^2)= 1.60(2)$ in good agreement
with what is expected phenomenologically. The ratio 
$2C_5^A(Q^2)/G_A(Q^2)= 1.63(1)$
is also independent of $Q^2$. Equality of these two ratios implies the 
Goldberger-Treiman relations. 
We also studied the $Q^2$-dependence of the form factors separately.
Comparing quenched and unquenched results at
pion mass of about 360~MeV, we observe large unquenching effects
on the low $Q^2$-dependence of the four form factors, $G_A(Q^2), \>G_P(Q^2), 
\>C_5^A(Q^2)$ and $C_6^A(Q^2)$. This
confirms the expectation that pion cloud effects are expected
to be large at low $Q^2$. Further study of pion
cloud effects using 
lighter quark masses on a finer lattice is called for. 

\vspace*{0.4cm}
\noindent
 {\bf Acknowledgments:}
A.T.  acknowledges support by the University of Cyprus and 
the program ``Pythagoras'' of the Greek Ministry of Education.
This work is
supported in part by the  EU Integrated Infrastructure Initiative
Hadron Physics (I3HP) under contract RII3-CT-2004-506078 and by the
U.S. Department of Energy Office of Nuclear Physics under contract 
DE-FG02-94ER40818. Computer resources were provided by NERSC under  contract No. DE-AC03-76SF00098.
\normalsize

\end{document}